# Wavefront shaping for opaque cylindrical lenses

Diego Di Battista[1,2], Haisu Zhang[1], Daniele Ancora[1], Krystalia Lemonaki[1], Evangelos Liapis[1], Stelios Tzortzakis[1,2,3], Giannis Zacharakis[1]

1-Institute of Electronic Structure and Laser, Foundation for Research and Technology-Hellas, N. Plastira 100, Vasilika Vouton, 70013, Heraklion, Crete, Greece.
2- Materials Science and Technology Department, University of Crete, 71003, Heraklion, Greece
3- Science Program, Texas A&M University at Qatar, P.O. Box 23874 Doha, Qatar

**Wavefront shaping has revolutionized the concepts of optical imaging and focusing. Contrary to what was believed, strong scattering in the optical paths can be exploited in favor of light focusing through turbid media and ultimately improve optical imaging and light manipulation capabilities. The use of light shapers and appropriately engineered scattering structures, i.e. opaque lenses enables the production of nano-scale confined foci and of extended fields of view. Exploiting this concept we fabricate configurable scattering structures by direct femtosecond laser writing. The properly shaped light trespassing the customized structure – a photonic lattice of parallel rods – forms a light-sheet at user defined positions. We demonstrate that our technique enables light-sheets with sub-micron resolution and extended depth of focus, a significant advantage when compared to the existing free space systems. Moreover, our approach permits to focus light of different wavelengths onto the same defined position without moving any optical element and correcting for chromatic aberrations. Furthermore, we demonstrate how an opaque cylindrical lens can be created using a biological structure, hence unveiling the potential advantages for optical microscopic and mesoscopic imaging.**

It is with immense interest that the scientific community follows the technological advancements of optical and photonic imaging and the creation of new and disruptive knowledge. It is in fact very recent that a Nobel prize was awarded to overcoming one of optics' most fundamental laws, the diffraction limit set by Abbe [1] and the creation of a new field. Nanoscopy or super resolution microscopy and the related techniques (STED, PALM, STORM) [2-6] are set to probe for the first time with optical wavelengths the structure of molecules instead of only their function. Furthermore, the technology and associated methods have moved from the laboratory bench to bedside in a rapid and ever increasing pace and have changed the paradigm of biological and biomedical imaging [7]. Methods that provide three dimensional microscopic images such as Optical Projection Tomography (OPT) [8] and light sheet microcopy (LSM) [9] or Selective Plane Illumination Microscopy (SPIM) [10] have found their way into biology labs in increasing numbers due to their advantages compared to traditional methods such as confocal microscopy [11]. Non-linear methods such as Second and Third Harmonic Generation (SHG and THG) provide non-labeling structural as well as functional information complimentary to the linear techniques [12]. These developments have provided the means for performing high resolution, quantitative, volumetric and dynamic studies in live specimens ranging from imaging the development [13], to imaging cancer [14], the function of the cardiovascular system [15], to neuroimaging, aging and associated diseases [16] to chemotherapeutic interventions and drug delivery [17]. Moreover, they offer to users unique capabilities for basic studies and most importantly for detecting and treating major diseases in clinical practice alongside or replacing established methodologies [18].

This has been achieved by a continuous development of the involved technologies, and the advent of innovative approaches exploiting prior knowledge from other fields of applied optics as well as new inventions in the effort to overcome the fundamental limits of optical imaging set by the very nature of light-matter interactions [7,12]. The main drawback of optical modalities originates from the multiple scattering that light suffers when propagating through complex media such as biological tissue. This strong scattering leads to image distortion and impaired image quality and resolution in the same way that fog lays a curtain between our eyes and sun light, blurring our vision [7,19].

Drawing knowledge from the seemingly distant field of astronomy, novel concepts of adaptive optics can be exploited to reverse image distortion, improve image resolution or retrieve otherwise hidden features [20-22]. In this context, wavefront shaping is one of the most efficient methodologies in compensating for scattering. By controlling the phase of the wavefront impinging on a scattering medium one can control the energy density of light at the output [23]. Wavefront shaping can always generate a focus at the back or inside complex systems if a feedback (guide-star) is provided [24], and is currently used for improving the imaging quality in deep biological tissues [25].

However, wavefront shaping covers a significantly wider range of applications or fields apart from biomedical imaging; manipulating the light transmitted through a complex medium has been demonstrated to allow communication control [26], to improve cryptography [27], to tune non-linear systems [28], to generate polarized foci [29] and reconfigurable Bessel beams [30]. However, the breakthrough that paved the way for wavefront shaping concerned optical focusing using a combination of a Spatial Light Modulator and a scattering system, which is also referred to in the literature as opaque lenses (OL) [31]. OLs can generate single and multiple foci at user defined positions by exploiting the speckle correlations, rendering in practice the focal position reconfigurable [23,31]. In addition, OLs exhibit a larger effective numerical aperture than conventional lenses, achieving, thus, resolutions comparable to those of super-resolution techniques [31-33] and extended Field of View (FOV) [34]. In practice, the focus has the same size as the speckle pattern correlation function at the beginning of the focusing process [31,35], and thus exploiting strong scattering samples high spatial frequencies are generated and super-confined foci are accessible [32-35].

In this work, we combine the two worlds of biomedical imaging and wavefront shaping to produce and manipulate improved illumination patterns for Light Sheet Microscopy by exploiting the properties of novel Opaque Lenses. We demonstrate that Opaque Lenses provide significant advantages in producing appropriate reconfigurable light sheets when compared to conventional cylindrical lenses. Cylindrical lenses focus light in only one dimension perpendicular to the direction of propagation, in such a way that they form a 2D light structure, the light sheet [9,10,13,14].

In LSM these light sheets are used to excite fluorescence from a section of a large biological sample hence performing real time optical sectioning and 3D imaging. A key challenge in LSM is the ability to generate a thin light sheet that maintains the geometrical characteristics along its propagation. Currently, the most promising efforts are based on free space beam shaping to produce non-diffracting beams, such as Bessel and Airy beams, which can be beneficial in certain configurations [36,37].

Here, we propose a method for generating reconfigurable light sheet illumination through laser printed scattering systems [38,39]. We use direct laser-writing to inscribe a dense photonic lattice consisting of scattering parallel rods randomly distributed inside the bulk of a thin silica glass plate. When light passes through the photonic lattice a light pattern composed of elongated speckles is produced. By introducing an appropriate phase mask, using a SLM we can enhance the intensity of one of these structured speckle grains generating a tightly confined light sheet. Our intent is to exploit the major effective numerical aperture of opaque lenses for generating super thin light sheets. Our system consisting of a SLM, a focusing spherical lens and the photonic lattice will henceforth be referred to as an Opaque Cylindrical Lens (OCL). Our approach enables the production of thin multi-color light sheets at user controlled positions without the need for mechanical scanning, exhibiting better axial resolution and extended field of view when compared to presently used approaches [36].

The unique properties of OCLs can significantly impact if not revolutionize applications of microscopic and mesoscopic optical systems as well as other fields of optics, enabling deep, fast and structured illumination and promoting real time, high resolution and quantitative imaging of live specimens.

## Results

**Opaque cylindrical lens.** When a coherent light beam travels through opaque structures experiences multiple scattering and at the output of the structure the light is fully scrambled with components presenting arbitrary k-vectors and phases. The superposition of these components generates a speckle-like interference pattern which consists of a set of intensity maxima and minima distributed in a complex pattern. A speckle pattern has been demonstrated to be complex rather than random which means that it conserves memory of the scattering structure [40] from which it was generated. In our case the scattering media are composed of randomly distributed rods piled along one direction in such a way to result parallel to each other (see Figure 1(a)). Such a structure presents a strong symmetry in one dimension that can be reflected on the speckle pattern generated. The samples are fabricated by direct laser writing (see Methods section), a schematic of the fabrication process is illustrated in Figure 1(a) and 1(b) and the final structure imaged with a 15X microscope objective is reported in Figure 1(c).

The interference pattern generated onto the camera plane placed at the back of the structures (Figure 2(a)) is composed of elongated speckle grains oriented with the same direction as the rods in the structures. This observation was confirmed by calculating the speckle pattern correlation function along the *x* and *y* directions of the camera plane: Figure 2(b) shows a comparison between the two. The profile of the correlation function provides the typical speckle size. The ratio between the Full Width at Half Maximum (FWHM) of the profile along *y* to those along *x* gives the *factor of speckle elongation* (FOSE), which in our case amounts to the value FOSE ≈ 3 (we measured FWHM = 0.9µm along the *x* axis and FWHM = 2.85µm along the *y* axis). This confirms that the typical speckle grain is indeed elongated along the *y* axis.

In such a configuration a phase only Spatial Light Modulator (SLM) can shape the wavefront and modulate light traveling through the synthetic scattering media. This approach was first presented in 2007 [41], using a layer of opaque white airbrush paint that improved the effective numerical aperture of the focusing system with respect to free space propagation. Recently, it has also been demonstrated that the presence of strong scattering allows opaque lenses to confine light in regions smaller than the diffraction limit and resolve sub-100 nm structures with visible light [32-34].

Herein we present a similar comparison; in our experiment a cylindrical lens ($f_1$=150mm) generates a light sheet at its focus, while a long working-distance microscopy objective (Numerical Aperture 0.65) and an eyepiece (200mm focal length) produce a magnified (40X) image on the CCD camera plane (schematic representation in Figure 3). We consider the Full Width at Half Maximum (FWHM) of the focus intensity profile as the focus width $w$ which determines the nominal focusing resolution. In free space we obtain a light-sheet with a lateral resolution of $w_{0X}$=6.7µm (see Figure 3.a). Calculated as the Full Width at Half Maximum of the intensity profile of the sheet along the direction $x$, i.e. its width. When we insert our synthetic scattering structure, which is the case in Figure 3.b, we maintain the same geometry but we add, 148.9mm ($f_{2A}$=148.9mm) at the back of the cylindrical lens, the 2D disordered structure. Using an iterative algorithm we are able to find the phase mask that allows generating a light sheet 0.1mm ($f_{2B}$=0.1mm) at the back of the structure. In this case we obtain a lateral resolution of $W_X$=0.9µm (see Figure 3.b), a value which is the result of an average on 10 foci obtained at different positions on the camera (corresponding to different target positions). Thus, we have achieved an experimental effective diffraction limit for the light sheet with a consequent enhancing factor [31] of $w_X/w_{0X}$ = 0.13.

**Phase mask.** The phase mask we address to the SLM is composed of parallel stripes/columns, and at each band corresponds a de-phasing that depends on the gray tone set on it. It follows that the beam impinging on the cylindrical lens is modulated along one dimension only. When the stripes of the phase mask are parallel to the rods of the structure the speckle pattern at the output persists in having elongated grains as shown in Figure 2.a and 3.b and the light sheet foci are available. On the contrary when segmented masks (see the example in Figure 3.c) are used, the geometry of the system is broken and the foci achieved at the end of the optimization process are confined in the typical circular region as shown in Figure 3.c. Such optimized parallel phase mask takes advantage of the directionality of the system reducing the complexity of the focusing algorithm from $N^2$ pixels to $N$ stripes, resulting in a faster focusing process.

**Fast multi-wavelength light sheet with sub-micron resolution.** Light Sheet Microscopy requires fast scanning (ideally in real time) of the samples at different wavelengths. In order to correct for the chromatic aberration one has to slightly change the position of the lenses at the illumination to maintain the sample on focus for the different wavelengths. This introduces delays and uncertainties, which are not suitable for fast *in vivo* scanning of the subjects. In our approach, we focus the light 100µm behind the 2D structure and we use three laser sources emitting at 488nm, 532nm and 594nm. Running the focusing process we find the three phase masks which allow the light sheets to co-localize at the same position for the three different colors. In such a way, we can change the illumination wavelength by switching from one mask to the other without the need of moving the position of lenses or of inserting shutters or filters along the light path. In Figures 4a, 4b and 4c we show the three light sheets obtained onto the same target position. For this we have maintained the same geometry shown in Figure 3.b only replacing the cylindrical with a spherical lens allowing a more precise control of the shaped beam. This provides the speckle pattern correlation function at the back of the sample as shown in Figure 2 (the average speckle elongation FOSE is maintained).

We then consider the intensity profiles of the final foci to obtain the average lateral resolution of $\overline{w}_X$=0.9µm along the *x* axis (see Figure 4d) and $\overline{w}_y$=21µm along *y* axis (see Figure 4e). For the calculation we create 10 light sheets at different times onto 10 different target positions and we average to extract the final resolutions. We repeat the same process for each wavelength and evaluate a change of

$\Delta w_X = 0.03 \mu m$ in the light sheet width passing from 488nm to 594nm wavelength. For applications with resolution around the micro-scale such small values of $\Delta w_X$ are insignificant.

We also measure the axial resolution $w_z$ of our light sheet scanning along $z$ the region around the plane where the focus was initially generated ($z_0$). We move the system composed of the objective lens, eyepiece and camera $100 \mu m$ in front and at the back of $z_0$ and we collect a frame each $5 \mu m$. By registering all the frames collected we can reconstruct the light sheet propagating along $z$ as shown in Figure 4d. We measure the axial resolution $w_z = 67 \mu m$ calculating the FWHM of the intensity profile along the direction $z$ as shown in Figure 4g.

One can easily find the optimal masks for focusing at defined positions for each wavelength, and thus is able to switch from one color to the other simply by addressing a different mask to the SLM. Considering that the modern SLMs are able to achieve refresh rates as high as 500Hz, the user can be equipped with a very fast tool that allows scanning the specimens at different wavelengths in a small fraction of a second.

**Focusing through Dentinal Tubules.** To test the approach described above on a biological sample, we chose histological Dentinal Tubule (DT) slices as the scattering medium. DTs are structured with parallel rods randomly distributed in the matrix of the teeth. This characteristic makes them remarkably similar to our synthetic structures composed of artificial rods. We used dry DT ground sections, such as those presented in the micrographs in Figure 5.a. As shown in panel a, the tubules run parallel to each other forming a 2D structure. Using thicker slices a structured scattering medium is formed (see Figure 5.b). When we impinge onto it with a coherent beam an interference pattern composed of elongated grains is generated at the back of the slice, as in the example reported in Figure 5.c. Calculating the autocorrelation from the image in Figure 5.c we can estimate the average orientation of the speckle grains, in the direction along which they present the elongation. The autocorrelation results with a central pick being elongated toward a direction which exhibits an angle $\vartheta$ with respect to the vertical axis of the camera. Therefore, we also rotate the SLM mask at the same angle $\vartheta$ in order to correctly align the phase bands parallel to the tubules direction (see Figure 5.d). A misalignment of the parallel phase mask with respect to the directionality of the tubules produces a break to the intrinsic dimensionality of the system and leads the convergence of the focusing algorithm to a conventional focal spot (not elongated). Once we select a target on the camera plane and run the focusing process we are able to produce a light sheet at the back of the DTs, as shown in Figure 5e, with a width of $w_x = 1.8 \mu m$ and a length of $w_y = 17.5 \mu m$.

## Discussion

As mentioned above opaque lenses produce sub-diffraction limit resolution [32-34] and extended Field of View [34] in comparison with the conventional optical systems in free space. With the introduction of our opaque cylindrical lenses we expect not only to look deeper in tissues but also to resolve nanoscale structures of large specimens. The opaque cylindrical lenses we studied are able to improve the resolution of conventional cylindrical lenses at a given focal length by a factor 8. The light sheet lateral resolution achieved is comparable with those presented in most recent works on free space beam shaping in the field of light sheet microscopy [36]. Nevertheless, the axial resolution that we obtain is slightly longer than the ones presented in the same works resulting in more uniform illumination deeper in turbid media.

Moreover, for the first time in the field of wavefront shaping a complex scattering medium is fabricated *ad hoc* to produce structured foci, in this case light sheets. Furthermore, our technique can be effectively scaled: for achieving, for instance, higher resolutions one can reduce the diameter of the photonic lattice rods. Lithographic techniques could push this limit quite low and significantly outperform present techniques [38].

In addition, with our technique we can easily correct for chromatic aberrations as we demonstrated above by generating light sheets at different wavelengths on the same spatial position without moving any of the system components. In practice, for a given focal length the light sheets produced at different wavelengths will slightly differ in shape due to the difference in refractive index experienced through the structure, but they will be generated on the same plane giving the possibility to switch from one color to the other by only changing the phase mask addressed to the SLM.

We would like to note that similar results have been demonstrated implementing metasurfaces that are able to focus the light and correct for chromatic aberrations [42, 43]. Our system presents a significant advantage when compared with the reported metasurfaces which resides in the fact that it is completely configurable; to change the focus position or the wavelength the user simply needs to change the mask addressed to the SLM. Moreover, 2D structures very similar to our artificial photonic lattice can be found in nature, e.g. muscle or collagen fibers, dental enamel and others. Probing these structures with polarized light can disclose information on the molecular organization of the investigated specimen. For this reason, they are lately subject of studies in both the linear and non-linear regimes [44,45].

Particular biological structures have already been demonstrated to produce interesting sub-diffractive light features [46]. To confirm and validate, as a proof of principle, the robustness and usability of our method in biological systems, we generated light sheets through dentinal tubule slices in arbitrary but fully controlled orientations depending on the orientation of the tubules within the slice [47]. These results demonstrate the potential of our approach in exploiting biological opaque lenses and improving the imaging capabilities and configurability of our systems.

## Methods

**Opaque cylindrical lens fabrication:** The scattering photonic lattice sample is fabricated by direct femtosecond laser writing in the bulk of a soda-lime glass plate (a microscope slide with dimensions 76x26x1$mm^3$). The high power femtosecond laser pulses provided by a Ti:Sapphire laser system, with a central wavelength of 800nm, a Fourier-transform-limited duration of 35fs, and a repetition rate of 1kHz, are tightly focused into the glass through the NA=0.75, 40x microscope objective. The incident pulse energy is set to be 8$\mu J$ (measured before the objective), which is high enough to induce micro-explosion and void-formation deep inside the soda-lime glass. Then the glass is scanned across the laser focus by motorized stages, in order to create the 3D structure. The lattice structure starts from 100$\mu m$ and extends to 600$\mu m$ underneath the glass surface, consisting of 100 layers of void-lines and each layer containing 70 void-lines located in random x-positions (see Fig. 1(a)). A typical void line has a width of 2$\mu m$ and a length of 200$\mu m$, which is inscribed by moving the glass straight along the y-direction at a scan speed of 700$\mu m/s$. The adjacent layers are 5$\mu m$ apart, and the minimum distance between adjacent void-lines in the same layer is 2$\mu m$ (see Fig. 1(b)). A top-view optical transmission image of the fabricated lattice structure is shown in Fig. 1(c). In essence, the structure consists of an ensemble of parallel rods randomly distributed in the volume of the glass , which form a scattering medium with a refractive index mismatch of ∆n=0.45 between the glass matrix and the void rods.

**Wavefront shaping setup:** A coherent laser source emitting at 594nm is used while a homemade telescope (lens L1 with 25.4mm focal length + lens L2 with 250mm focal length) magnifies the laser beam by 10X. Modulation is performed by a phase only Spatial Light Modulator (SLM) (Holoeye, Pluto) that shapes the wavefront of the beam; a 50:50 beam splitter (BS) guarantees the beam and the SLM are perpendicular to each other. Hence, 50% of the light reflected by the SLM is directed along a perpendicular axis where a cylindrical or spherical lens focuses the beam at its focal distance. The beam impinges onto the scattering sample (S) and the output is collected by a 40X infinity corrected microscope objective (OBJ) which projects the sample output onto the camera plane.

**Iterative algorithm:** The focusing process is achieved exploiting an iterative Monte-Carlo algorithm similar to those presented in previous works [48]. A mask composed of 200x200 segments is addressed to the SLM window and we randomly assign to each segment of the first line a gray tone on a scale of 255 levels corresponding to a fixed de-phasing of the light reflected by the same segment within a range from 0 up to $4.5\pi$. The same sequence is repeated for all the lines underlying the first. The result is a mask composed of *N*=200 columns with different gray tones as shown in Figure 3.b. A preliminary optimization step tests a series of 50 random masks picking the one providing the best intensity value at the target position. The mask is used as an input for the second optimization routine, which tests a phase shift from 0 to $2\pi$ for a single column accepting the one providing the highest enhancement measured at the target. Subsequently, each column of the mask is tested.

**Dentinal Tubules Slices:** DTs are microscopic sigmoid ('S') shaped curved channels which contain the long cytoplasmic processes of odontoblasts and extend radially from the dentinoenamel junction (DEJ) in the crown area, or dentinocemental junction (DCJ) in the root area, to the outer wall of the pulp, forming a network for the diffusion of nutrients throughout dentin [49]. Tubules run horizontally from the inside of the tooth to the outside located within the dentin and near the root tip, incisal edges and cusps, the dentinal tubules are almost straight. Gradually narrowing from the inner to the outermost surface of the teeth, the tubules have a diameter of 2.5µm near the pulp, 1.2µm in the middle of the dentin, and 0.9µm at the dentin-enamel junction. Their density is 59,000 to 76,000 per square millimeter near the pulp, whereas the density is only half as much near the enamel [50]. In that respect, dentinal tubule slices resemble and can be used as natural 2D scattering structures. In our study we used a 0.7mm thick dentinal tubule ground section (DENT-EQ, Basavanagudi, Bangalore, India), enough to scramble completely the impinging light and generate a complex speckle pattern with negligible ballistic contribution.

## Acknowledgements

This work was supported by the GSRT Aristeia projects "Skin-DOCTor" (1778) and "FTERA" (2570) as well as the "Neureka!" "Supporting Postdoctoral Researchers" project of the "OPERATIONAL PROGRAMME EDUCATION AND LIFELONG LEARNING", co-funded by the European Social Fund (ESF) and National Resources. This work was also supported from the EU Marie Curie Initial Training Network "OILTEBIA" PITN-GA--2012-317526, the FP7-REGPOT CCQCN (EC-GA 316165) and the H2020 Laserlab Europe (EC-GA 654148).

## Author Contributions

All authors have contributed to the development and/or implementation of the concept. D.D.B., G.Z., S.T. conceived and designed the experiment. D.D.B. and K.L. realized the experimental apparatus and performed the measurements. H.Z. fabricated and characterized the 2D structures. D.A. analyzed the data and supported the development of the experiment.  E.L. provided and measured the dentinal tubules slices.

# Figures

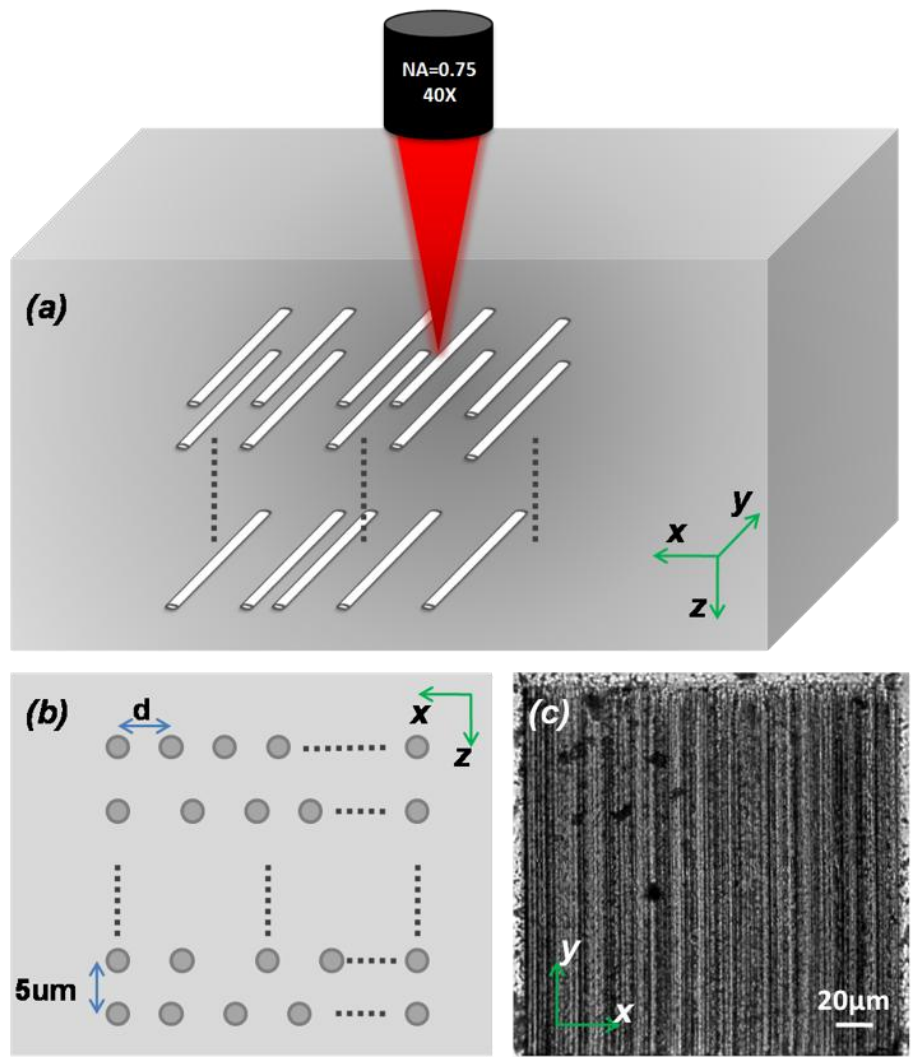

*Figure 1:* Fabrication details. The scattering rod photonic lattice is fabricated in the bulk of the glass by direct laser writing as illustrated in panel (a). The distribution of the lattice structure on the x-z plane is depicted in panel (b). An optical transmission micrograph of a fabricated rod lattice is shown in panel (c).

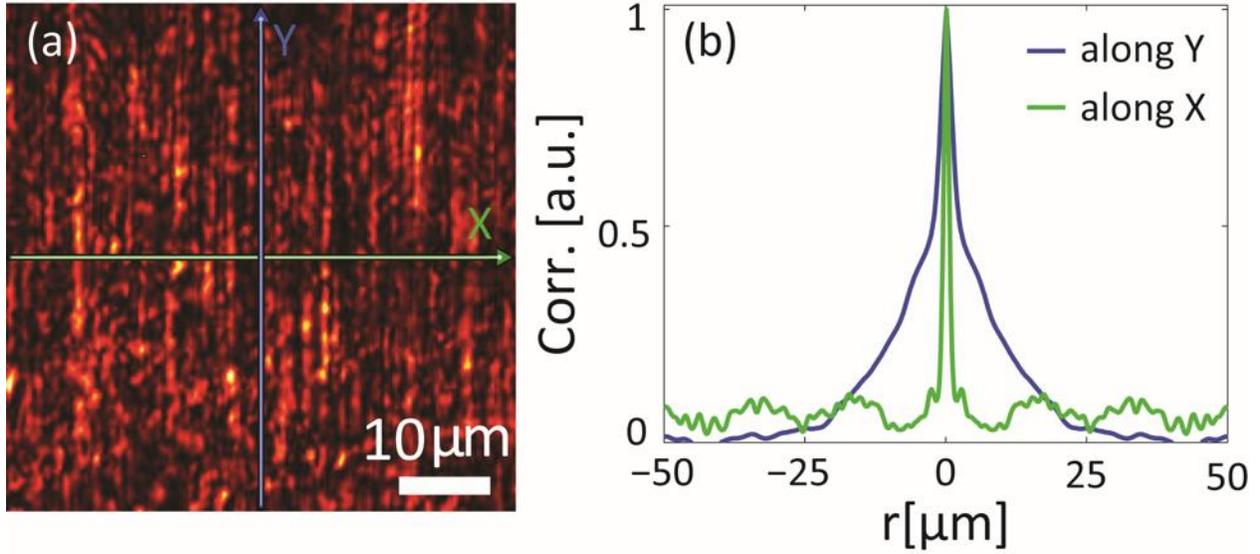

*Figure 2:* *The speckle pattern at the back of the photonic lattice: the isotropy along the y axis in the structure results in a speckle pattern with elongated speckle grains (panel a). On the panel (b) we extrapolated the typical speckle grain size calculating the correlation function of the pattern along the direction x (green solid line) and the direction y (blue solid line). Comparing the full width at half maximum of the two profiles the curve along y resulted approximately 3 time larger than that along x.*

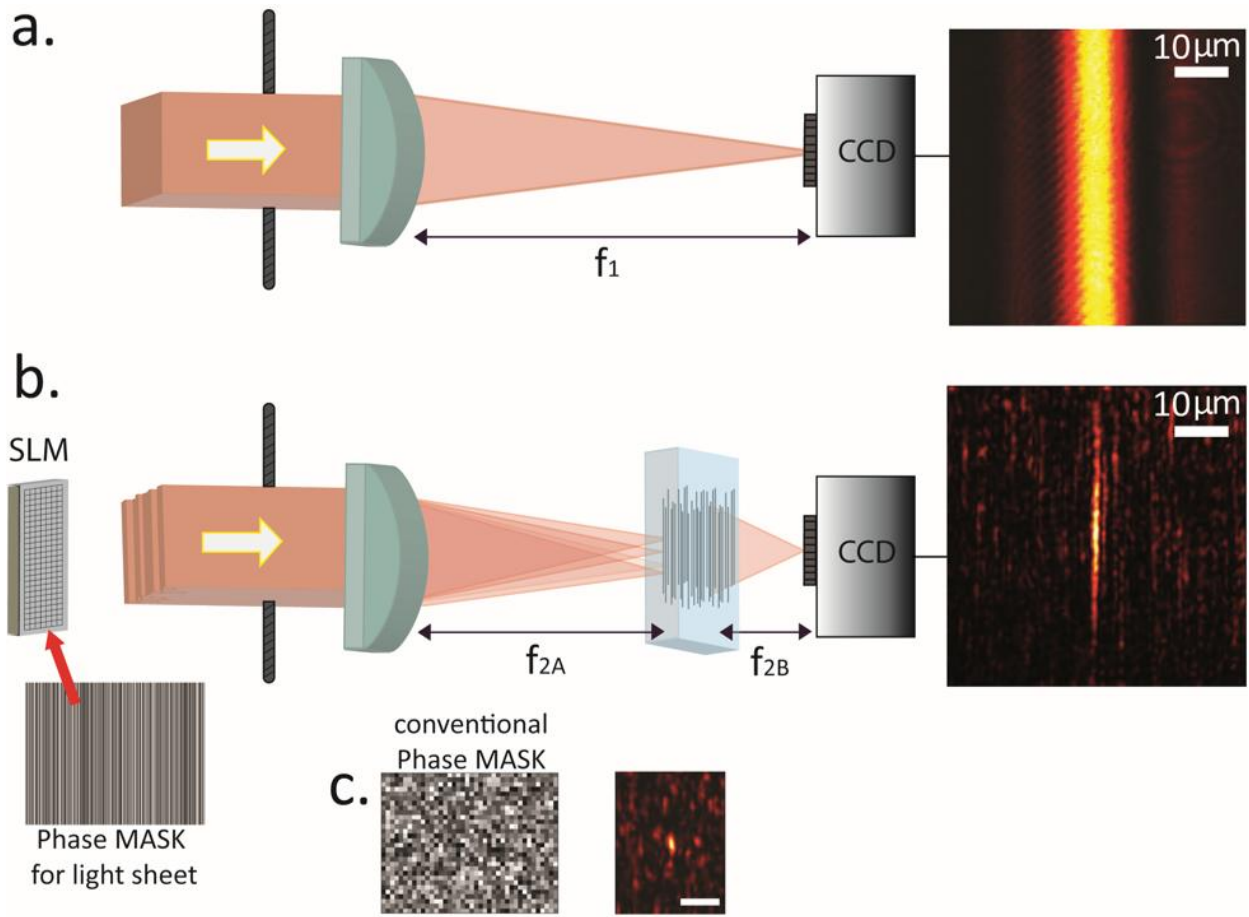

*Figure 3:* *A cylindrical lens generates a light sheet at its focus $f_1$=150mm, two pinholes control the aperture D of the lens. On panel a. we show the free space configuration. On b. we introduce the 2D photonic lattice 148.9 mm in front the lens. We shape the beam addressing to the SLM phase masks composed of parallel stripes in order to conserve the system isotropy. In such a way we enhance the elongated grain on the target position generating the light-sheet. If segmented masks as the one on the left of panel c. are used for the optimization process the 2D geometry is lost and a conventional round focus (illustrated on the right side of panel c.) is generated.*

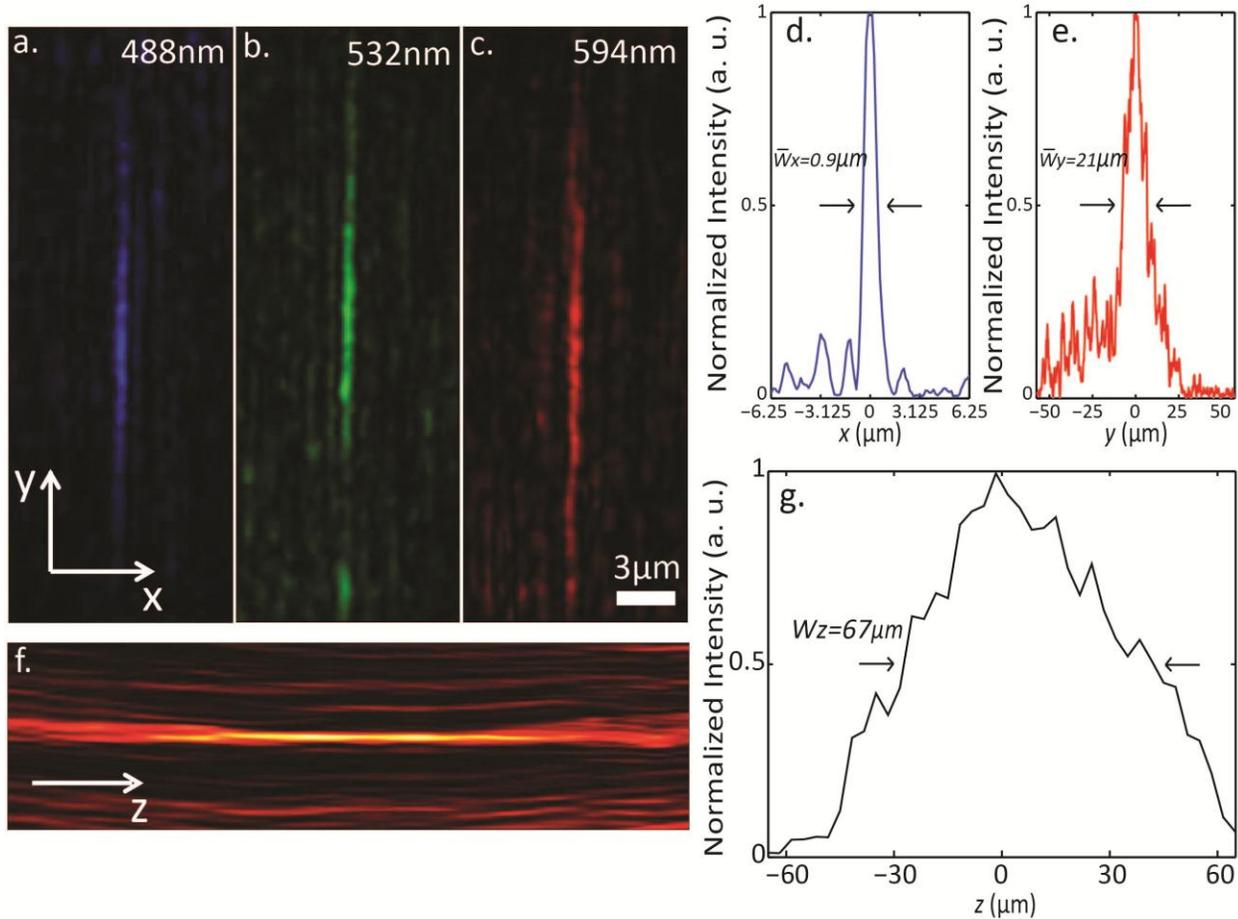

*Figure 4:* *Images a., b. and c. show respectively light-sheets generated at 488nm, 532nm and 594nm laser source wavelengths. Measuring the FWHM of the intensity profile along the x axis (blue curve in panel d.) and the FWHM of the intensity profile along the y axis (red curve in panel e.) we extrapolated the light-sheet width ($\bar{w}_x = 0.9\mu m$) and length ($\bar{w}_y = 21\mu m$). In (f.) we show the light-sheet along the z axis. To evaluate the effective depth of the focus generated we measured the FWHM of the intensity profile along z ($w_z = 67\mu m$) as shown in plot g.*

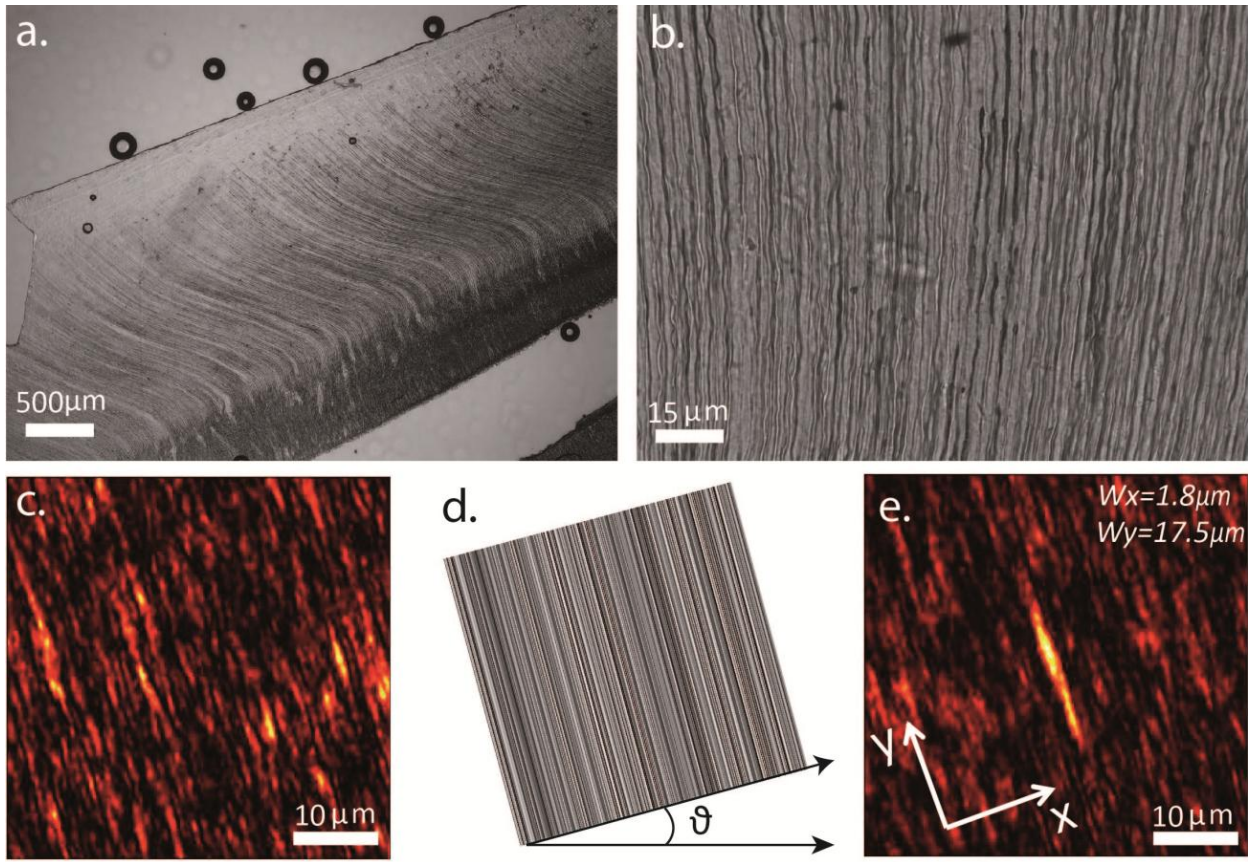

*Figure 5:* *Adaptive focusing through a Dentinal tubule slice (DTS). On image (a.) a thin DTS shows parallel tubules running along the dentinal section. In b. a thick DTS forms a perfect 2D structure. When the light passes through such a curtain forms an interference pattern composed of elongated speckle grains isotropically oriented along one direction given by the tubules direction as shown in c. θ is the angular distance between the orientation of the grains and the vertical camera axis. In order to maintain the isotropy in one dimension we rotate the SLM window at an angle θ as illustrated in the sketch d. The stripes of the phase mask will be parallel to the dentinal tubules. Sketch e. is the light-sheet at the end of the focusing process.*